\documentclass[a4paper,fleqn,usenatbib]{mn2e}
\usepackage{graphicx}
\usepackage{amsmath}
\usepackage{amssymb}
\usepackage{upgreek}

\title[Radio light curve of FRB\,150418]
{Radio light curve of the galaxy possibly associated with FRB\,150418}

\author[Johnston et al.]  {S.\,Johnston$^1$\thanks{email: simon.johnston@csiro.au},
E.\,F.\,Keane$^{2,3,4}$,
S.\,Bhandari$^{3,5}$,
J.-P.\,Macquart$^{6}$,
S.\,J.\,Tingay$^{6,7}$,\newauthor
E.\,Barr$^{8}$,
C.\,G.\, Bassa$^{9}$,
R.\,Beswick$^{4}$,
M.\,Burgay$^{10}$,
P.\,Chandra$^{11}$,
M.\,Honma$^{12,13}$,\newauthor
M.\,Kramer$^{8,4}$,
E.\,Petroff$^{9}$,
A.\,Possenti$^{10}$,
B.\,W.\,Stappers$^{4}$,
H.\,Sugai$^{14}$
\\
$^{1}$CSIRO Astronomy and Space Science, Australia Telescope National Facility, PO Box 76, Epping, NSW 1710, Australia\\
$^{2}$Square Kilometre Array Organisation, Jodrell Bank Observatory, SK11 9DL, UK\\
$^{3}$Centre for Astrophysics and Supercomputing, Swinburne University of Technology, Mail H29, PO Box 218, Victoria 3122, Australia\\
$^{4}$Jodrell Bank Centre for Astrophysics, School of Physics and Astronomy, University of Manchester, Manchester M13 9PL, UK\\
$^{5}$Australian Research Council Centre of Excellence for All-sky Astrophysics (CAASTRO), Australia\\
$^{6}$International Centre for Radio Astronomy Research (ICRAR), Curtin University, Bentley, WA 6102, Australia\\
$^{7}$Istituto Nazionale di Astrofisica (INAF) -- Istituto di Radioastronomia, Via Pietro Gobetti, Bologna, 40129, Italy\\
$^{8}$Max-Planck-Institut f\"ur Radioastronomie (MPIfR), Auf dem H\"ugel 69, D-53121 Bonn, Germany\\
$^{9}$ASTRON, the Netherlands Institute for Radio Astronomy, Postbus 2, NL-7990 AA Dwingeloo, The Netherlands\\
$^{10}$Istituto Nazionale di Astrofisica (INAF)-Osservatorio Astronomico di Cagliari, Via della Scienza 5, I-09047 Selargius (CA), Italy\\
$^{11}$National Centre for Radio Astrophysics, Tata Institute of Fundamental Research, Pune University Campus, Ganeshkhind, Pune 411 007, India\\
$^{12}$Department of Astronomical Science, SOKENDAI (Graduate University for the Advanced Study), Osawa, Mitaka 181-8588, Japan\\
$^{13}$National Astronomical Observatory of Japan, 2 Chome-21-1 Osawa, Mitaka, Tokyo 181-8588, Japan\\
$^{14}$Kavli Institute for the Physics and Mathematics of the Universe (WPI), Institutes for Advanced Study, University of Tokyo, Kashiwa, Chiba 277-8583, Japan\\
}

\date{Accepted \today. Received \today; in original form \today}

\pubyear{2016}

\begin{document}
\label{firstpage}
\pagerange{\pageref{firstpage}--\pageref{lastpage}} 
\maketitle

\begin{abstract}
We present observations made with the Australia Telescope Compact Array (ATCA),
the Jansky Very Large Array (JVLA) and the Giant Metre-Wave Telescope of
the radio source within the galaxy WISE~J071634.59$-$190039.2, claimed to be 
host of FRB~150418 by Keane et al. (2016). We have established a 
common flux density scale between the ATCA and JVLA observations, the
main result of which is to increase the flux densities obtained by Keane et al.
At a frequency of 5.5~GHz, the source has a mean flux density of 140~$\upmu$Jy
and is variable on short timescales with a modulation index of 0.36.
Statistical analysis of the flux densities shows that the variations seen are
consistent with refractive interstellar scintillation of the weak 
active galactic nucleus at the centre of the galaxy.
It may therefore be the case that
the FRB and the galaxy are not associated. However, 
taking into account the rarity of highly variable sources in the radio sky,
and our lack of knowledge of the progenitors of FRBs as
a class, the association between WISE~J071634.59$-$190039.2 
and FRB~150418 remains a possibility.
\end{abstract}


\section{Introduction}
Fast Radio Bursts (FRBs) are a relatively new class of astrophysical
phenomenon, the origin of which remains an open question. Although the
Parkes radio telescope has detected the majority of the FRBs
(e.g. \citealt{lbm+07,kkl+11,tsb+13,bsb14,rsj15,pjk+15,cpk+16,kjb+16}),
others have been detected at Arecibo \citep{sch+14}
and at the Green Bank Telescope \citep{mls+15}.
Observationally, FRBs are millisecond-duration bursts of radio emission
with high dispersion measures (DMs), some of which appear to be
polarized. Some FRBs show evidence of scattering, others are 
temporally unresolved within observational limits while yet others
have a double-peaked structure \citep{cpk+16}. FRB~121102 shows multiple
bursts with different temporal and spectral features 
\citep{ssh+16a,ssh+16b} whereas extensive observations of other
FRBs have failed to see repetition \citep{lbm+07,pjk+15}.

The DMs of the FRBs greatly exceed the DM contribution expected from
the Milky Way by factors of as much as 30; this has led to the 
interpretation that the FRBs are extragalactic \citep{lbm+07}.
In addition, observations of the scattering seen in some FRBs lends credence 
to the extragalactic interpretation \citep{mj15,cws+16} whereas the Galactic 
interpretation (e.g. \citealt{lsm14}) seems barely tenable \citep{den14}.
While the extragalactic interpretation of FRBs thus seems reasonably secure,
the same cannot be said of their progenitors with many ideas
currently in the literature (e.g. \citealt{fr14,lyu14,cw16,kat16}).
Indeed, given the observational phenomenology listed above, it remains unclear 
whether all FRBs have the same origin.
If FRBs are indeed extragalactic in origin and can be detected in 
large numbers at cosmologically interesting distances, they can then be 
used as tools for high precision cosmology \citep{mcquinn14,mkg+15}.
For example, Keane et al. (2016)\nocite{kjb+16} used the DM of FRB~150418
in conjunction with the optical redshift of the putative associated
galaxy to measure the cosmic density of ionized baryons, and found the result
to be in agreement with that of WMAP \citep{hlk+13}.

The FRBs in the literature to date have all been discovered with
single dish telescopes and hence have localisation of only a few arcmin
at best. This means
there are many sources at both optical and radio wavelengths within
the positional uncertainty of the FRB, making any putative association
difficult. Progress in determining the progenitors (or afterglows) of
FRBs depends on obtaining an accurate position and although valiant
attempts have been made to detect FRBs with interferometers over a wide
range of frequencies
(e.g. \citealt{wbd+11,sbf+12,cvh+14,kca+15,lbb+15,ttw+15,rbm+16,cfb+16,btb+16}),
arcsecond positions remain elusive.

The SUrvey for Pulsars and Extragalactic Radio Bursts (SUPERB) detected
FRB~150418 in real time using the Parkes radio telescope \citep{kjb+16}.
Follow-up observations using the Australia Telescope Compact Array (ATCA) and 
the Subaru optical telescope led to \citet{kjb+16} claiming that a 
fading radio transient was associated with the galaxy 
WISE~J071634.59$-$190039.2 at redshift $z=0.49$.
The low probability of finding such a radio source within the
field of view of the Parkes telescope \citep{bhh+15,mhb+16} meant that
\citet{kjb+16} were able to argue that the transient was associated with the 
FRB and hence located in the optical galaxy.

Although initially classified as a {\it transient}, data taken 
subsequent to the \citet{kjb+16} paper made it clear that the source
is more correctly described as a {\it variable}. This led to
the association between the FRB and the radio transient being called
into question by \citet{wb16}, who claimed the variability was 
intrinsic to an active galactic nucleus (AGN), and \citet{aj16} who 
proposed the variability was caused by refractive scintillation of the AGN.
Jansky Very large Array (JVLA) observations by \citet{vrm+16}
covering a wide range of frequencies, showed that radio emission
from WISE J071634.59-190039.2 had flux densities comparable to the
later epochs of \citet{kjb+16}. The source of the emission was
shown to be compact on EVN baselines \citep{gmg+16}. VLBA and
e-MERLIN radio observations and optical imaging with Subaru \citep{bbt+16}
showed that the compact radio source is located in the center of 
WISE J071634.59-190039.2, and suggest that the source is likely a weak AGN.

In this paper we describe observations of the radio source performed
with the ATCA, the JVLA and the Giant Metre Wave Telescope (GMRT) in Section 2.
Section 3 presents the results with special attention given to the 
absolute flux density calibration between the ATCA and JVLA observations.
Section 4 is given to interpretation of the radio light curve, before we
present our conclusions in Section 5.

\section{Observations and Data Reduction}
\begin{table*}
\caption{Observations and flux density of the radio source associated with 
WISE~J071634.59$-$190039.2.}
\label{tab:flux}
\begin{tabular}{lllllcllrl}
\hline
Date & Start & Length & MJD & Telescope & Band & Array & Mode & Flux Density & Error \\
& UT & (hr) & & & (GHz) & & & \multicolumn{2}{c}{($\upmu$Jy)}\\
\hline \hline
2015 Apr 18 & 06:30 & 5.5 & 57130.271 & ATCA & 4.5 -- 6.5 & 6A & Mosaic & 320 & 18\\
            &       &       &           &      & 6.5 -- 8.5 & 6A & Mosaic & 225 & 27\\
2015 Apr 24 & 02:44 & 12.0 & 57136.114 & ATCA & 4.5 -- 6.5 & 6A & Mosaic & 154 & 19\\
            &       &       &           &      & 6.5 -- 8.5 & 6A & Mosaic & $<$90 & 30\\
2015 Apr 26 & 01:45 & 10.0 & 57138.073 & ATCA & 4.5 -- 6.5 & 6A & Mosaic & 106 & 17\\
            &       &       &           &      & 6.5 -- 8.5 & 6A & Mosaic & $<$90 & 30\\
2015 May 18 & 12:30 & 2.0  & 57159.521 & GMRT & 0.69 -- 0.72 & & Single & $<$350 \\
2015 May 22 & 12:42 & 2.0  & 57163.529 & GMRT & 1.37 -- 1.40 & & Single & $<$150 \\
2015 Jul 04 & 21:12 & 7.4 & 57208.884 & ATCA & 4.5 -- 6.5 & 6D & Mosaic & 98 & 17\\
            &       &       &           &      & 6.5 -- 8.5 & 6D & Mosaic & $<$90 & 30\\
2015 Oct 27 & 14:09 & 8.5 & 57322.59  & ATCA & 4.5 -- 6.5 & 6D & Mosaic & 98 & 17\\
            &       &       &           &      & 6.5 -- 8.5 & 6D & Mosaic & $<$75 & 25 \\
2015 Nov 11 & 18:40 & 8.5 & 57336.778 & GMRT & 1.37 -- 1.40 & & Single & $<$70 \\
2016 Feb 24 & 06:09 & 7.6 & 57442.4   & ATCA & 4.5 -- 6.5 & 6B & Mosaic & 106 & 13\\
            &       &       &           &      & 6.5 -- 8.5 & 6B & Mosaic & $<$75 & 25\\
2016 Mar 01 & 04:18 & 1.2 & 57448.17  & JVLA  & 4.0 -- 8.0 & C  & Mosaic & 132 & 10\\
2016 Mar 01 & 05:29 & 11.0 & 57448.5   & ATCA & 4.5 -- 6.5 & 6B & Single & 121 & 6\\
            &       &       &           &      & 6.5 -- 8.5 & 6B & Single & 111 & 8\\
2016 Mar 10 & 10:29 & 5.0 & 57457.5   & ATCA & 4.5 -- 6.5 & 6B & Single & 143 & 6\\
            &       &       &           &      & 6.5 -- 8.5 & 6B & Single & 59 & 8\\
2016 Mar 11 & 10:25 & 5.0 & 57458.5   & ATCA & 4.5 -- 6.5 & 6B & Single & 140 & 6\\
            &       &       &           &      & 6.5 -- 8.5 & 6B & Single & 225 & 11\\
2016 Mar 13 & 10:23 & 5.0 & 57460.5   & ATCA & 4.5 -- 6.5 & 6B & Single & 137 & 6\\
            &       &       &           &      & 6.5 -- 8.5 & 6B & Single & 156 & 11\\
\end{tabular}
\end{table*}
\begin{table*}
\caption{Flux density at 5.5~GHz of the radio source associated with 
WISE~J071634.59$-$190039.2 from observations taken on the JVLA by \citet{wb16}.
For 2016 March 23 we use their published value rather than the ATEL value.}
\label{tab:wb}
\begin{tabular}{llllrlrl}
\hline
Date & Start & Length & MJD & Flux Density & Error & Flux Density & Error \\
& UT & (hr) & & \multicolumn{2}{c}{($\upmu$Jy)} & \multicolumn{2}{c}{($\upmu$Jy)} \\
& & & & \multicolumn{2}{c}{(Williams \& Berger)} & \multicolumn{2}{c}{(this paper)}\\
\hline \hline
2016 Feb 27 & 00:25 & 1.5  & 57445.03 & 156 & 11 & 169 & 17\\
2016 Feb 28 & 00:11 & 1.5  & 57446.02 & 153 & 13 & 165 & 17\\
2016 Mar 05 & 00:07 & 0.5  & 57452.02 & 105 & 21 & 116 & 14\\
2016 Mar 08 & 23:54 & 0.5  & 57456.01 & 225 & 24 & 192 & 17\\
2016 Mar 11 & 23:29 & 0.5  & 57458.99 & 147 & 26 & 151 & 14\\
2016 Mar 16 & 23:04 & 0.5  & 57463.97 & 279 & 25 & 254 & 15\\
2016 Mar 23 & 22:58 & 0.5  & 57470.97 & 218 & 24 & 153 & 16\\
2016 Mar 28 & 22:52 & 0.5  & 57475.97 & 259 & 21 & 221 & 16\\
2016 Mar 31 & 23:12 & 0.5  & 57478.98 & 205 & 15 & 189 & 11\\
2016 Apr 01 & 22:00 & 0.5  & 57479.93 & 260 & 34 & 270 & 28\\
\end{tabular}
\end{table*}
Observations of the field containing FRB~150418 were made with the ATCA, 
the GMRT and the JVLA covering a period of close to one year after the FRB.
In addition, JVLA observations were made by \citet{wb16} and
\citet{vrm+16} following the publication of \citet{kjb+16}.
A series of observations were also made with the VLBA and e-MERLIN, 
described in detail in \citet{bbt+16} and with the EVN \citep{gmg+16}.

A total of 10 epochs were obtained with the ATCA of which the
first five were included in Keane et al. (2016).
Their observations plus those of
epoch 6 were made by mosaicing together 42 separate pointings.
For epochs 7 to 10 inclusive, a single pointing was used centered on the 
position of the WISE galaxy. All these observations were carried out
in two bands each of 2-GHz bandwidth, the first centered at
5.5~GHz and the second at 7.5~GHz. Flux calibration was
carried out with the ATCA calibrator B0823$-$500 while phase
calibration was performed using B0733$-$174.
ATCA data reduction was carried out using the {\sc miriad} package
\citep{stw95} using standard techniques. For the mosaiced observations, each
pointing was reduced separately and the pointings later combined
using the task {\sc linmos}. Each 2-GHz band was reduced independently.
Flux measurements were made using the task {\sc imfit} from the images 
and {\sc uvfit} directly from the calibrated {\sc uv} data.

Three epochs were obtained with the GMRT. The first and second epochs
used 3C286 as the flux calibrator whereas the third epoch used 3C147.
Phase calibration was carried out using 0837$-$198 for the first epoch and 
0735$-$175 for the last two epochs.
GMRT data reduction was carried out using {\sc AIPS}, the images were
then ported into {\sc miriad} for analysis.

A single epoch was obtained with the JVLA. A seven point mosaic was
used to tile the field of view surrounding the FRB. A total of 4~GHz
of bandwidth was used centered at 6.0~GHz.
Flux calibration was carried out with 3C147 and phase calibration with
B0733$-$174, the same phase calibrator as used for the ATCA.
Interference flagging and imaging were carried out using {\sc casa}.

Finally, the data taken by \citet{wb16} on the JVLA became publicly available
and were re-analysed. These observations were made with a single
pointing in two separate bands each with 1~GHz bandwidth centered at
5.5 and 7.5~GHz. Again, 3C147 and B0733$-$174 were used for flux and
phase calibration. Data reduction was carried out using {\sc casa}.
The images were ported into {\sc miriad} and flux density measurements were made
using the task {\sc imfit} in an identical fashion to the ATCA data.

In this paper, flux densities were obtained using {\sc imfit} under
the assumption that the source is unresolved at the
observing frequencies and baseline lengths of the ATCA, GMRT and JVLA,
an assumption justified by the long baseline results \citep{bbt+16}.

\subsection{Absolute flux density calibration}
The ATCA data described in Keane et al. (2016) and those subsequently obtained
on the ATCA were flux density calibrated using B0823$-$500, as the source was 
in angular promixity to the FRB field, whereas the primary ATCA flux 
density calibrator B1934$-$638 was not.
We did, however, also observe B1934$-$638 on most occasions.
B0823$-$500 is known to have intrinsic variability on yearly
timescales and indeed its flux density
has been increasing over the last decade making the flux density
model built into {\sc miriad} out of date (J. Stevens, private
communication). To quantify
this effect, we separately measured the flux density of the phase calibrator
B0733$-$174 using B0823$-$500 or B1934$-$638 as the flux calibrator.
We found that at 5.5~GHz, use of B0823$-$500 underestimates flux density 
measurements by on average 8\% whereas at 7.5~GHz this difference is 15\%.
We see only a marginal increase in the flux density of B0823$-$500 over the 
entire data span.
The flux densities described in Keane et al. (2016) therefore, 
although self-consistent, are too low by these amounts for 
the purposes of absolute comparison with other instruments.

In addition to this effect we have uncovered an issue with the way
that {\sc miriad} deals with the primary beam correction for
mosaiced images over wide bandwidths such as used in these observations.
The consequence of this effect is that the flux densities as measured
with mosaiced images are lower than those made with a single pointing.
To quantify this effect we took advantage of other sources within
the field of view of the large mosaic. We compared the flux densities
as measured with the mosaic with those measured when dealing solely
with the single pointing containing the source.
We find that the flux density of the single pointings are 10 percent
higher than those in the mosaics, because of the way {\sc miriad} handles
large fractional bandwidth in the {\sc linmos} routine.
A similar effect also appears to be present in the {\sc casa} reduction
of JVLA mosaics and affects the flux density that we made in
our mosaiced JVLA observation. We therefore also extracted the single pointing
of the JVLA mosaic that contained the WISE galaxy and processed 
these data independently.
\begin{figure*}
\begin{tabular}{cc}
\includegraphics[width=8cm]{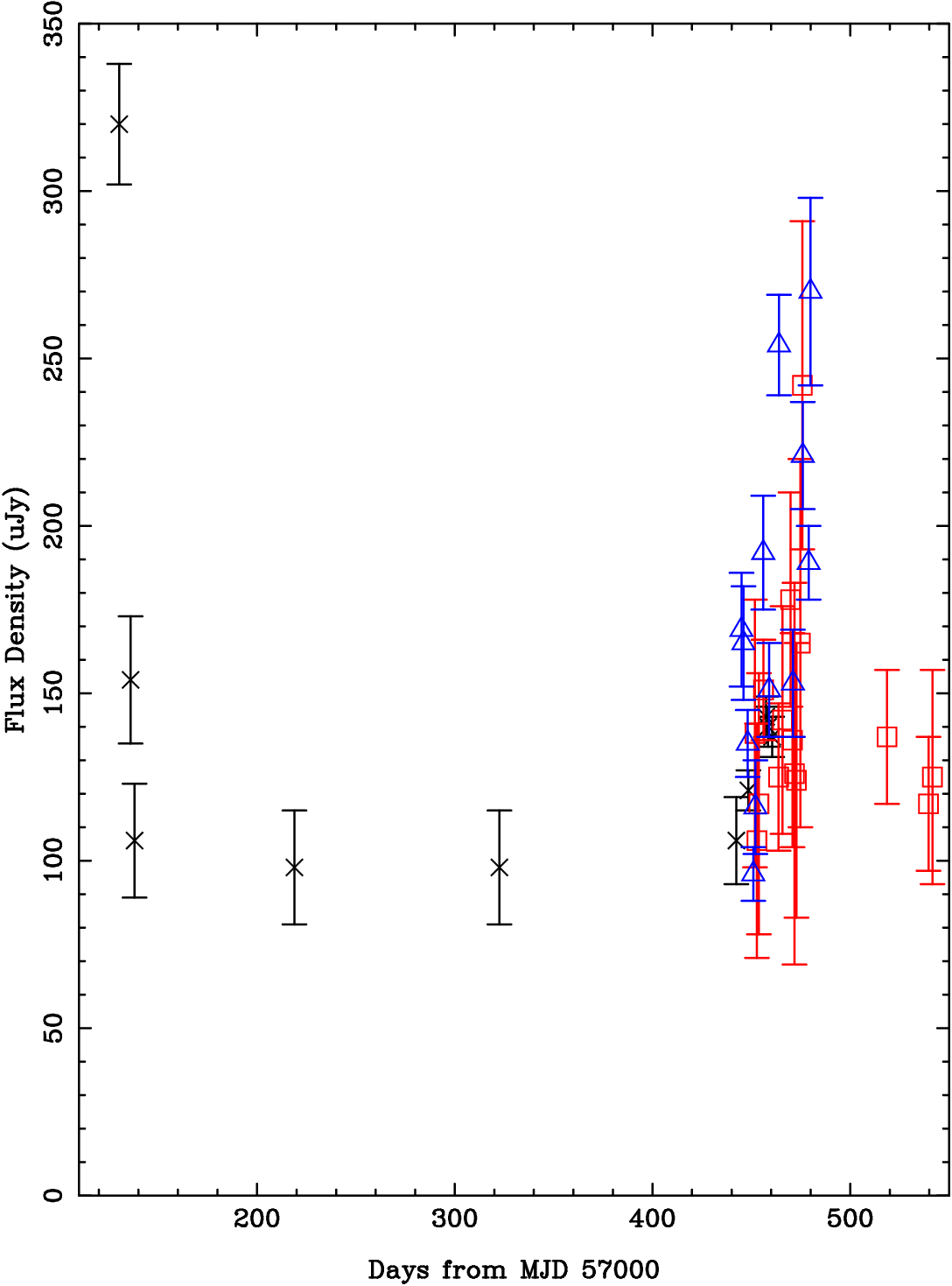} &
\includegraphics[width=8cm]{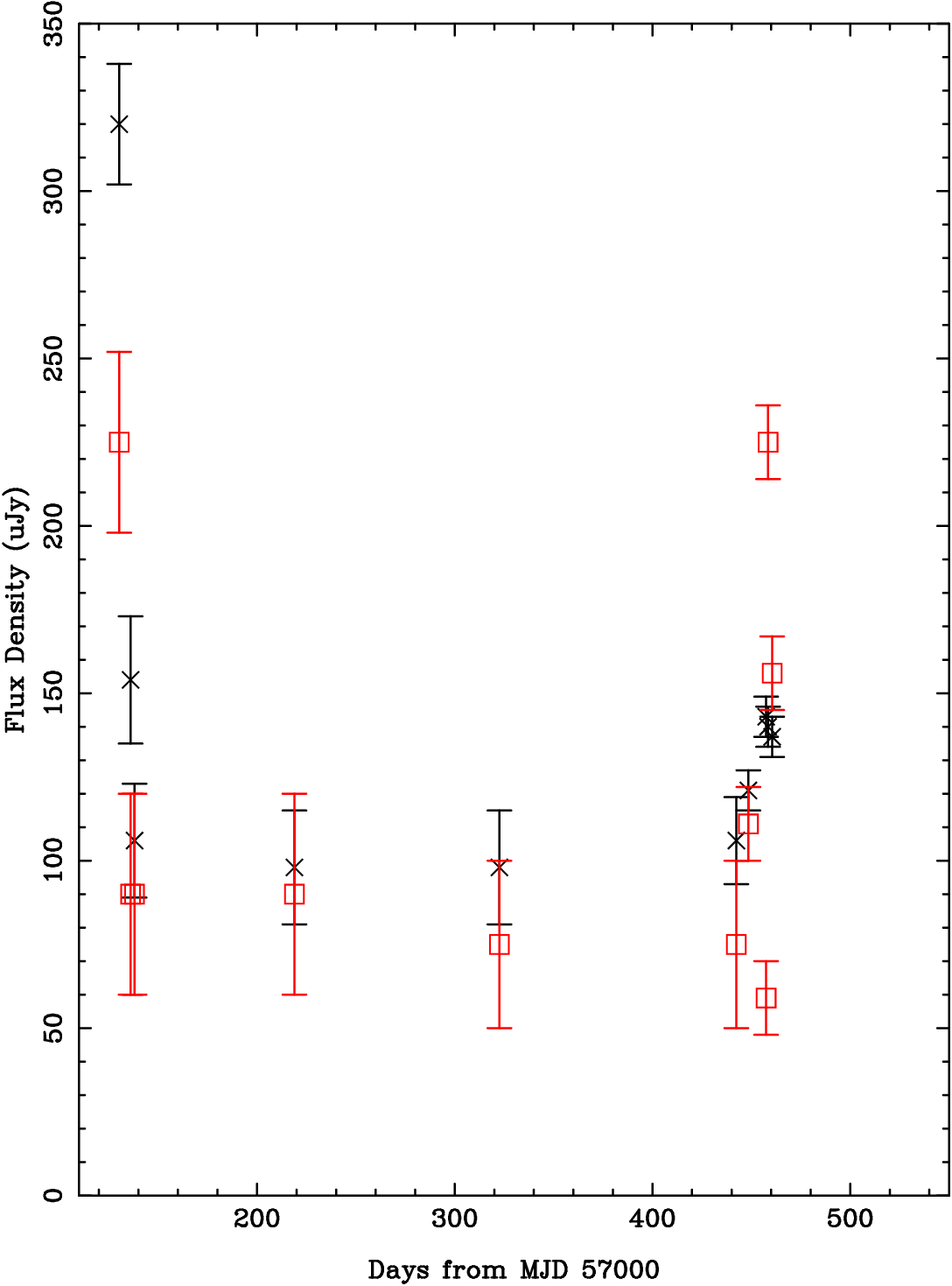} \\
\end{tabular}
\caption{Left panel: Radio light curve at 5.5~GHz. Black crosses denote ATCA 
observations, blue triangles JVLA observations, and red squares VLBA, e-MERLIN
and EVN.
Right panel: Radio light curve from the ATCA data at 5.5~GHz (black crosses) 
and 7.5~GHz (red squares).}
\label{fig:flux}
\end{figure*}

For the ATCA data, the two bands (from 4.5 to 6.5~GHz and from 
6.5 to 8.5~GHz) are treated independently. We have verified the calibration
between the two bands by examining the spectrum of the phase calibrator to
ensure that it forms a smooth power law over the entire 4~GHz range.

Finally in order to compare the JVLA and ATCA observations, we need to
ensure that there is overlap in the frequency coverage. The JVLA 
observation taken by us from 4.0 to 8.0~GHz is offset from the
ATCA observations from 4.5 to 8.5~GHz. The JVLA observations of
\citet{wb16} were centered at the same frequencies as the ATCA but only
had 1~GHz of bandwidth per sub-band rather than the 2~GHz of the ATCA.
Given the spectral behaviour of this source, it is important to keep
these differences in mind when trying to compare flux density measurements
from the different observations.

\subsection{Summary of data calibration}
In summary, the flux density values presented in \citet{kjb+16} are
self-consistent between epochs but are, in absolute terms,
low for two reasons. First, {\sc miriad} has an incorrect model
for the (current) flux density of B0823$-$500 and secondly {\sc miriad} 
incorrectly accounts for primary beam correction across a wide bandwidth
for mosaiced images. The total of these effects means that the
\citet{kjb+16} flux densities are too low by about 20 percent in comparison
to those obtained with the JVLA. If we require accuracy at the greater than
10 percent level all these effects need to be accounted for.

Potential pitfalls of measuring flux density have been described
in great detail in \citet{rbo16}.
Effects include clean bias, interference flagging, imaging
algorithms used ({\sc miriad} versus {\sc casa}), the effects of
wide bands and the software used to compute flux densities in either
the image or the uv domain. These effects are likely to account for
most of the residual differences between measurements noted below.

\section{Results}
Table~\ref{tab:flux} lists the observations and the measurements of the
flux density of WISE~J071634.59$-$190039.2.
Note that the date of the 4th epoch of ATCA observations
is 2015 July 4 and not June 4, as reported in \citet{kjb+16}.
Where the source was not detected at 7.5~GHz we give a $3-\sigma$ upper
limit.  The source was not detected in any of the GMRT observations and 
we quote an upper limit likewise.

In Table~\ref{tab:wb} we compare the flux density obtained with the JVLA
data as reported in \citet{wb16} with those obtained through data reduction
carried out by us. These agree within the 1-sigma error bars apart
from the data taken on 2016 Mar 23. We note that there is a discrepancy
between the result in the published paper at this epoch
(218$\pm$24~$\upmu$Jy) and that given
in ATEL~\#8946 (185$\pm$18~$\upmu$Jy) with the latter value consistent with 
our value.  In subsequent analyses we use the values that we derived.

We note that the brightness obtained with the VLBA on 2016 March 8 of
151$\pm$15~$\upmu$Jy/bm is significantly lower than that obtained only
17~h later with the JVLA (192$\pm$17~$\upmu$Jy) and the EVN brightness 
on 2016 March 16 was 125$\pm$22~$\upmu$Jy/bm as opposed to the JVLA value 
of 254$\pm$15~$\upmu$Jy taken only a few hours later.
One possible interpretation is that the EVN and VLBA
are resolving the structure leading to a lower brightness. We do not
believe this to be the case, however, as the intermediate resolution
data from e-MERLIN are consistent with both the ATCA/JVLA and the EVN/VLBA
measurements.

\begin{figure}
\includegraphics[width=8cm]{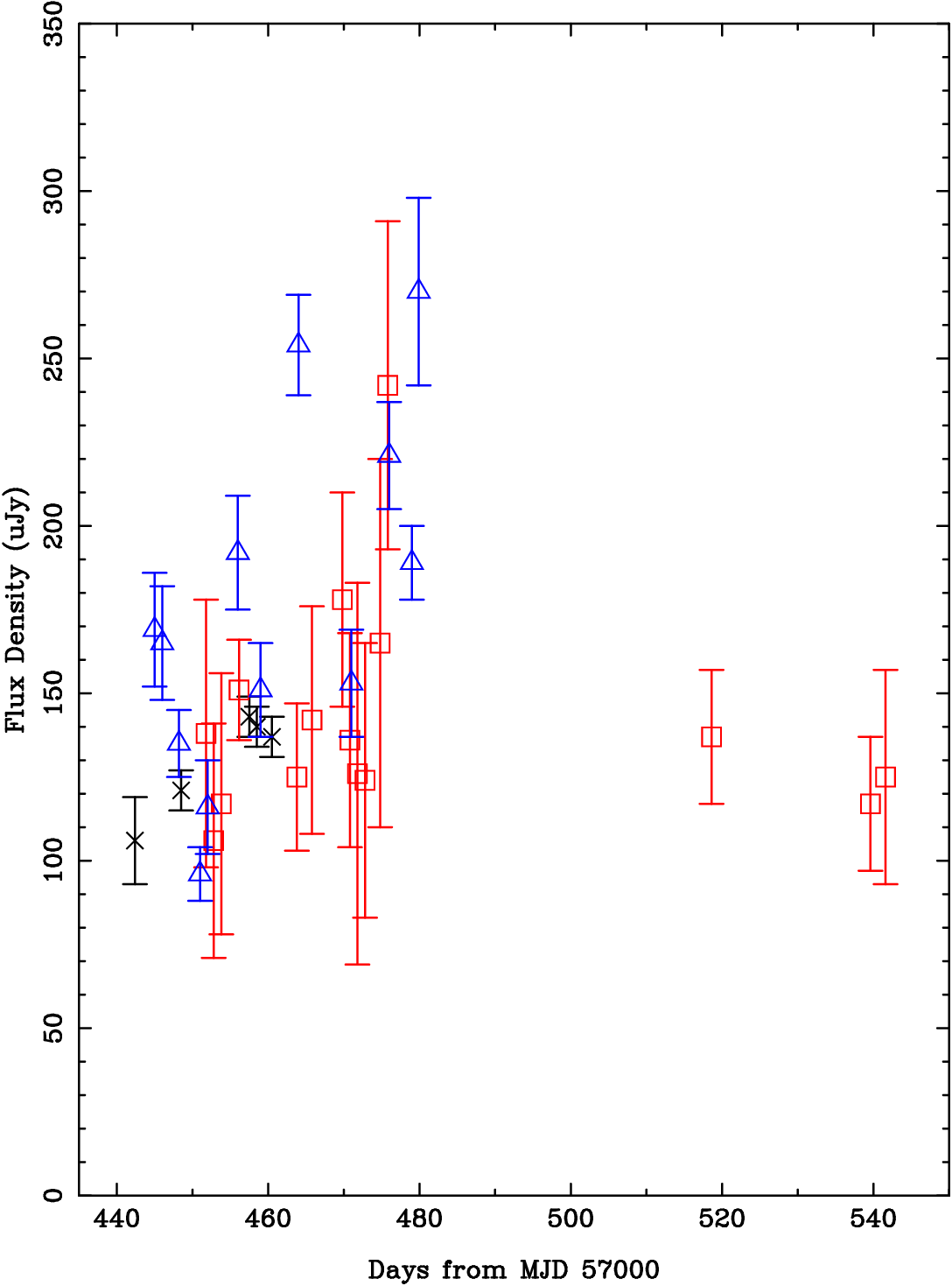}
\caption{End portion of the radio light curve at 5.5~GHz. Black crosses denote 
ATCA, blue triangles JVLA and red squares VLBA, EVN or e-MERLIN.}
\label{fig:zoom}
\end{figure}

The left hand panel of Figure~\ref{fig:flux} shows the light curve at 
5.5~GHz.  Red points are values obtained with the ATCA, blue points
from the JVLA, and green points from VLBA, e-MERLIN \citep{bbt+16}
and the EVN \citep{gmg+16}.
The right hand panel of the figure shows the 5.5~GHz and 7.5~GHz light curves
from measurements taken on the ATCA.
It can clearly be seen that the first two epochs
have a strongly falling spectrum (negative spectral index), whereas for
epochs 3 to 7 the spectrum is flat.
Figure~\ref{fig:zoom} shows in more detail the final 100 days of
observations.

\begin{figure}
\includegraphics[width=8cm]{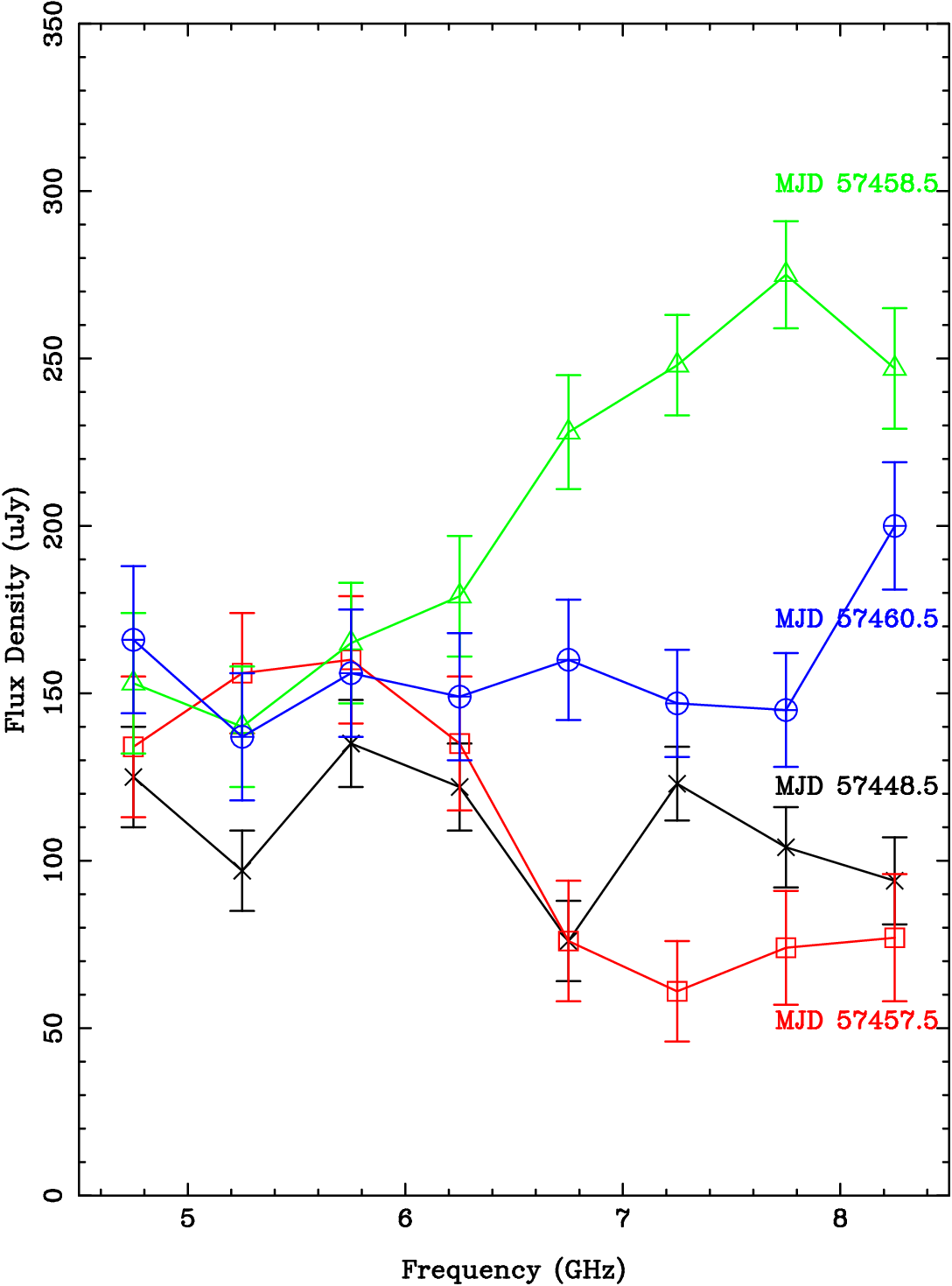}
\caption{Flux density as a function of frequency for MJDs 57448.5 (crosses
and black line), 57457.5 (squares and red line), 57458.5 (triangles and
green line) and 57460.5 (circles and blue line) of the ATCA data.}
\label{fig:spec}
\end{figure}

\begin{figure}
\includegraphics[width=8cm]{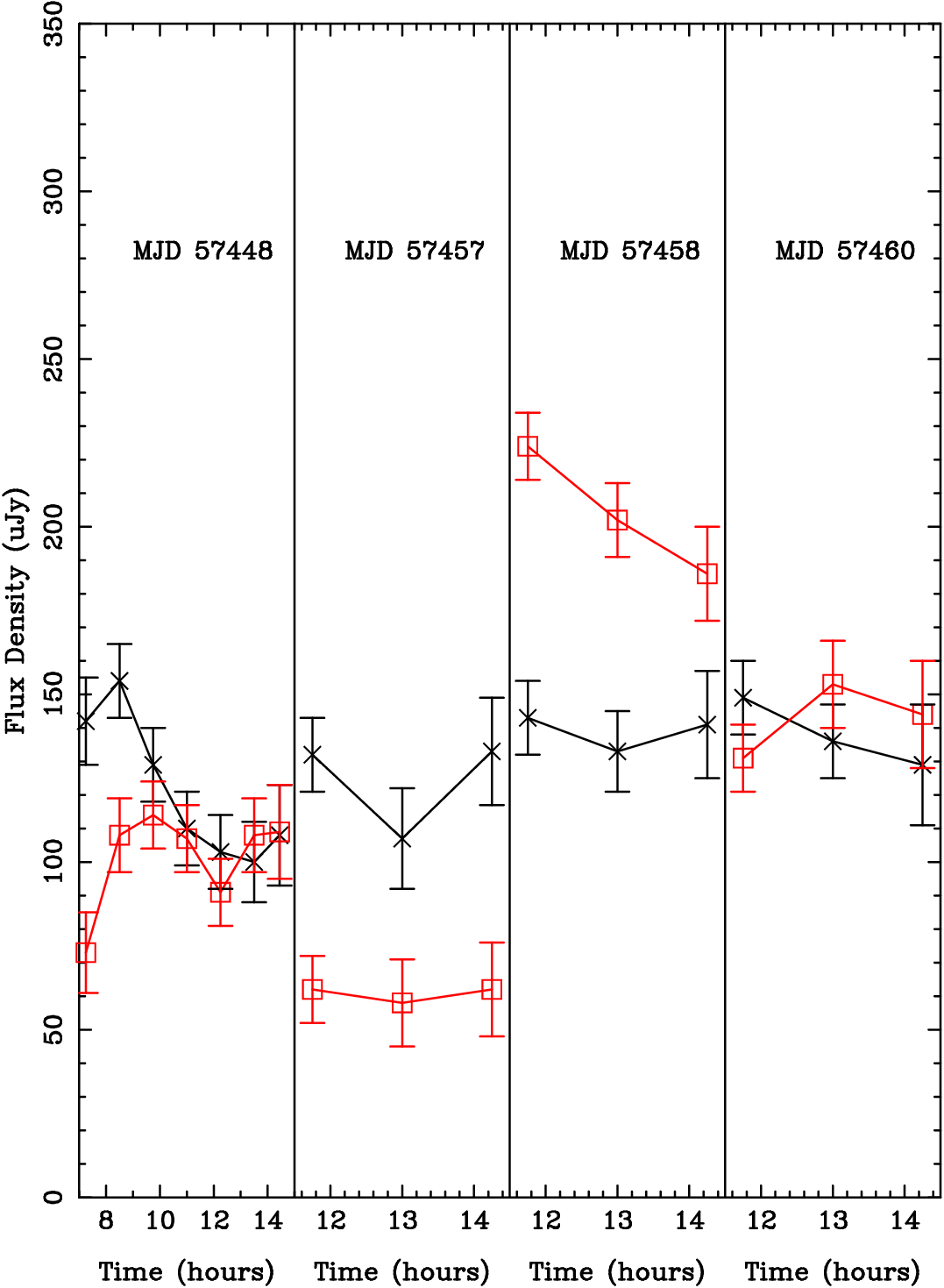}
\caption{Flux density as a function of time for epochs 7 through 10.
Each panel shows a different epoch with the UT time in hours from the
MJD listed at the top of the panel.
Points with crosses and black lines denote 5.5~GHz, 
points with squares and red lines denote 7.5~GHz.}
\label{fig:time}
\end{figure}

For the ATCA observations from epochs~7 to 10, where only single pointings
of long duration were obtained, the signal to noise ratio of the radio source 
is significantly higher than in the earlier epochs.  This allows the data 
to be subdivided in frequency and time bins.
Figure~\ref{fig:spec} shows the spectra for these 4 epochs.
The spectral index is highly variable.
Finally, Figure~\ref{fig:time} shows the flux density as a function of
time for epochs 7 to 10 at both 5.5 and 7.5~GHz. For each epoch we subdivide
the data into 2.5-h time averages in steps of 1.25~h. At 5.5~GHz there
is marginal evidence for variability on the timescale of a few hours
but nothing like the factor of two change between the measurements
on the JVLA and EVN on 2016 March 16. At 7.5~GHz significant time variability
is seen from epoch to epoch.

\section{Discussion}
\subsection{Variability}
For a formal metric for variability we use the
debiased modulation index, $m_d$, and $\chi^2$ statistic from \citet{bhh+15}
as well as the variability statistic, $V_S$, from \citet{mhb+16}.
These are defined by:
\begin{equation}
 m_{d} = \frac{1}{\overline{S}} \sqrt{\frac{\sum_{i=1}^{n}(S_{i} - \overline{S})^{2}   - \sum_{i=1}^{n} \sigma_{i}^{2}  }{n}}, 
\end{equation}
\begin{equation}
\chi_{lc}^{2} = \sum_{i=1}^{n} \frac{(S_{i} - \tilde{S})^{2}}{\sigma_{i}^{2}}.
\end{equation}
\begin{equation}
V_S = \frac{{\rm S_{max}} - {\rm S_{min}}}{\sigma}
\end{equation}
where $n$ is the number of data points, $S_i$ and $\sigma_i$ are the
flux density and the error bar for each datum, $\overline{S}$ and
$\tilde{S}$ are the mean and weighted mean flux densities, and
$S_{\rm max}$ and $S_{\rm min}$ are the
maximum and minimum flux densities recorded.

Combining the data taken at 5.5~GHz with the ATCA and the JVLA, 
$\tilde{S} = 140$~$\upmu$Jy, $m_d=0.36$, $\chi^2=250$ and $V_S=11$.
These values change only slightly if the 
e-MERLIN and VLBA points from \citet{bbt+16} are included.
We note that, in spite of our best efforts at determing the absolute
calibration, $\tilde{S}$ for the \citet{wb16} data 
(181~$\upmu$Jy) is significantly larger than the ATCA data.
The data taken at 7.5~GHz at later epochs yield
$\tilde{S} = 120$~$\upmu$Jy, $m_d=0.39$, $\chi^2=180$ and $V_S=12$.
Both the modulation indices and the variability timescales are in
line with those predicted by \citet{aj16} at these Galactic latitudes.

Observationally, a source with these varability metrics is rare and,
by these measures, both \citet{bhh+15} and \citet{mhb+16}
would clearly classify this source as a variable.
\citet{ofb+11} have only 12 out of 464 sources ($\sim$0.5 percent)
with $\chi^2$ greater than 100, of these only 2 have a modulation index in 
excess of 0.2. In the MASIV survey of \citet{lrm+08}, of their 443 sources 
specifically chosen to be flat spectrum and hence susceptible to
variability, only 1 source per epoch showed a modulation index in excess of
0.2. The survey of \citet{mhb+16} found that 38 out of 3652 sources
($\sim$1 percent) had modulation indices in excess of 0.26 over the period of
a week. The usual caveats apply here; all these surveys were either at a 
different observing frequency or a higher flux density limit or both,
compared to our data.

\subsection{Interstellar scintillation}
\citet{bbt+16} and \citet{gmg+16} show that the ATCA radio 
source (at least at later epochs) is highly compact and located at the 
centre of the optical galaxy.
This indicates that the source is likely an AGN.
Although the $\sim$mas resolution is not sufficient to determine
whether the source should undergo interstellar scintillation (ISS) it
is certainly indicative. In addition, the quasi-simultaneous observations
made by \citet{vrm+16} between 1 and 20~GHz indicate that the source has a 
flat spectral index (modulated by ISS; see \citealt{aj16}).

We deduce from the observations that the variations may be dominated by 
refractive scintillation. Diffractive scintillation appears unlikely to 
be contributing to the flux density variations: at the position of the source 
the expected timescale for diffractive scintillation is $\simeq 10\,$minutes,
for an assumed scintillation velocity of 50\,km\,s$^{-1}$, 
which is much shorter than the timescale of the fastest variations evident in 
the lightcurve. Moreover, we note that diffractive scintillation is only 
relevant if the source contains structure on scales below $\sim 0.1\upmu$as
for the expected scattering properties along this line of sight.   

We therefore investigated the probability that the data are consistent with the
form of an intensity probability distribution function expected from ISS
in the regime of refractive scintillation. This is a Rician
distribution \citep{cgh+67} with two free parameters, the amplitude of the 
non-varying component and the amplitude of the scintillation fluctuations.
The probability that the 5.5~GHz light curve is drawn from such a 
distribution is 90 percent.

Using simple arguments we would also expect that the modulation index
of weak radio sources be higher than those of brighter sources if all
AGN have a brightness temperature near the Inverse Compton limit. If
this is correct then the high modulation index we measure for
WISE~J071634.59$-$190039.2 compared to sources in e.g. the MASIV survey
\citep{lrm+08} is to be expected. In addition, the low galactic
latitude of WISE~J071634.59$-$190039.2 also likely contributes to the
higher modulation index.

We therefore conclude that the variability we see in
WISE~J071634.59$-$190039.2 is largely due to ISS with a couple of minor 
caveats. First, the probability distribution argument does not
take into account the time sequence of the data. Although we have no
{\em a priori} knowledge of what form an afterglow to an FRB might take,
the light curve from \citet{kjb+16} is suggestive of a fading radio 
transient. Indeed, even with the additional data, the probability
of getting a value as high as that of the first epoch is only 0.3 per cent.
Secondly, ISS in compact AGN is highly intermittent \citep{lrm+08} and
the light curve is only sparsely sampled during the first few months.

\subsection{Probability arguments}
The data obtained by \citet{kjb+16} appeared to show that the radio
source was a {\em transient} (as opposed to a variable).
As outlined in \citet{kjb+16}, comparison with the \citet{ofb+11}, 
\citet{mhb+16}, and \citet{bhh+15} surveys yield
probabilities of finding such a transient source at 
significantly less than 1 percent. \citet{lz16} also provided 
a compelling chain of logic to show that 
the probability of seeing the variability of WISE~J071634.59$-$190039.2 
is only 0.1 percent when using the \citet{ofb+11} survey as a comparison.
With the additional data we have obtained for WISE~J071634.59$-$190039.2,
it is clear that the source is more correctly
described as a {\em variable} and the values computed above
need to be re-visited. In particular the \citet{mhb+16} survey showed 
that some 1 percent of their sources have a modulation 
index in excess of 0.26 on timescales of less than a week.

What then is the probability of detecting a source undergoing ISS in the 
field of view of the ATCA? The ATCA image covers an area of 0.0625 sq
degrees and is roughly complete above 100~$\upmu$Jy;
the source counts at this observing frequency above this
flux density level are $\sim$300 per square degree \citep{hhl+12}.
At these flux density levels, star forming galaxies start to dominate the number
counts and these galaxies do not have strong compact components and will not
be seen as ISS variables at our sensitivity. \citet{hhl+12} estimate that 
30 percent of their sources have flat spectral indices consistent with 
AGN rather than star formation.
Finally, not all compact sources that could potentially undergo ISS are
actually doing so at any one epoch. \citet{lrm+08} estimate
the incidence of ISS at $\sim$20 percent, but the modulation indices
of their sources are only of order 0.1. Variability with a higher modulation 
index is rarer although, as noted above weaker sources likely modulate
more and WISE~J071634.59$-$190039.2 is closer to the Galactic plane than
the sources in MASIV.  We therefore estimate the incidence of ISS yielding
a modulation index in excess of 0.3 to be $\sim$5 percent at these
Galactic latitudes.
We are therefore left with an estimate of 1.5 percent of all sources,
or $\sim$4.5 sources per square degree above 100~$\upmu$Jy, are 
undergoing ISS at any one epoch. This estimate is not dissimilar to
the value given above obtained by consideration of the \cite{mhb+16} survey.

In summary, we estimate that of order 1 percent of all sources at 5.5~GHz
at a flux density limit of 100~$\upmu$Jy show variability with a high modulation
index on short timescales.  Within the positional error box of 
FRB~150418 we observe 8 sources brighter than 100~$\upmu$Jy and the 
probability of observing such a variable source is then $\sim$8 percent.

\subsection{The variable sky}
We have established that the fraction of variable sources, with parameters
at 5.5 GHz such as those for WISE~J071634.59$-$190039.2, 
is $\sim$1 percent, both from considering our knowledge of ISS
statistics and from the observed number counts of large-scale surveys.
Given 300 sources per square degree above 100~$\upmu$Jy at 5.5~GHz,
more than $10^5$ sources across the entire sky will show variability over 
the course of 1 week. This result holds irrespective of the association
or not with the FRB.

However, let us for a moment assume an afterglow interpretation for
the light curve over the first week. Then,
with $\sim$5000 FRBs per sky per day, any given week 
would then produce $3.5\times 10^4$ FRB afterglows. We therefore see no
obvious disconnect between the number of FRB afterglows and our current 
knowledge of the variable sky.
This is in contrast to \citet{vrm+16} who claimed that FRB afterglows
must be rare. There are three reasons for this difference. First,
\citet{vrm+16} assumed a modulation index in excess of 0.7 which new data
shows is too high. They also assumed the entire
flux density from the first epoch was the afterglow but clearly 
at least a significant fraction of the emission
originates in the underlying AGN. Finally, the volume for FRBs probed with
the Parkes telescope is significantly higher than the volume probed
for sources similar to WISE~J071634.59$-$190039.2 with the ATCA.
If indeed we are observing the afterglow to FRB~150418,
ATCA follow-up surveys would not detect them if they were significantly
more distant than $z\sim0.5$.

\section{Conclusions}
We have combined observations from the ATCA, GMRT, JVLA, eMERLIN, VLBA 
and the EVN to show the light curve for the radio source associated with
the galaxy WISE~J071634.59$-$190039.2.
The source, which Keane et al. (2016) associated with 
FRB~150418, has, with the acquisition of new data, been shown to 
be highly variable with a modulation index of 0.36,
a timescale for variability of order 1 day and a variable spectral
index. These statistics are consistent with those expected from ISS in
the weak scattering regime \citep{aj16} although the value seen in the first 
epoch, taken only 2 hours after FRB~150418, remains anomalously high.
We show that the probability of detecting such a scintillating source in the 
error box of the FRB is $\sim$8 percent, higher than
presented in Keane et al. (2016).

Given this information, two possibilities are open.
First, that the radio source is simply a weak AGN undergoing interstellar 
scintillation and is not associated with the FRB.
If this is the case, \citet{kjb+16} were unlucky with the low 
probability of detection, with the high flux density for the first epoch and 
the time sequence of the light curve. However, this option cannot be ruled out.
The second possibility is that the radio source is indeed associated with
the FRB. The FRB progenitor would then have to comply with the relatively 
low level of star formation in the galaxy \citep{kjb+16}. 
If this is typical for FRBs as a class, we might then 
expect to see a scintillating AGN in conjunction with an FRB event.
Only one other FRB has published follow-up to date \citep{pbb+15} but
only a single epoch of ATCA data was obtained.
Given the small numbers and our lack of understanding of the progenitors
of FRBs, the arguments for or against the association essentially depend on
the probability of observing such a strongly variable source within the
field of view of the FRB localisation.
Therefore, the \citet{kjb+16} conclusion is only one of several viable
conclusions, the relative probabilities of which are difficult to assess.

With only some 20 FRBs known to date \citep{pbj+16}, progress in the
field requires many more detections and accurate (arcsec) localisations
of the bursts themselves rather than relying on probability arguments
as here. By providing the latter, new instruments such as the
Canadian Hydrogren Intensity Mapping Experiment (CHIME;  \citealt{baa+14}),
the Hydrogen Intensity and Real-Time Analysis Experiment (HIREX;  \citealt{nbb+16}),
the Molonglo Observatory Synthesis Telescope (UTMOST; Bailes et al. In Prep),
the Aperture Tile in Focus on Westerbork (APERTIF; \citealt{vov+08}),
MeerTRAP (Stappers \& Kramer In Prep) and TRAPUM on
MeerKAT\footnote{http://www.ska.ac.za/science-engineering/meerkat/},
and the Commensal Real-time ASKAP Fast-Transient Survey (CRAFT; \citealt{mbb+10}) on the Australian Square Kilometre Array
Pathfinder (ASKAP; \citealt{dgb+09,jtb+09}) will undoubtedly 
break new ground.

\section*{Acknowledgments}
The Parkes telescope and the Australia Telescope Compact Array are part
of the Australia Telescope National Facility which is funded by the
Commonwealth of Australia for operation as a National Facility managed
by CSIRO. Parts of this research were conducted by the Australian
Research Council Centre of Excellence for All-sky Astrophysics (CAASTRO)
through project number CE110001020.
Parts of this work were performed on the gSTAR national
facility at Swinburne University of Technology. gSTAR is funded by
Swinburne and the Australian Government's Education Investment Fund.
CGB and EP acknowledge support from the European Research Council under the 
European Union's Seventh Framework Programme (FP/2007-2013) / ERC Grant 
Agreement nr. 337062 (DRAGNET; PI Hessels) and nr. 617199 respectively.
We thank Sarah Burke-Spolaor for assistance with the JVLA data reduction,
Matthew Bailes for stimulating discussions and the
EVN team for providing results prior to publication.
\bibliographystyle{mnras}
\bibliography{photo2}

\bsp
\label{lastpage}
\end{document}